\documentclass[prl,twocolumn,superscriptaddress]{revtex4}
\usepackage{graphicx}

\begin{document}

\title{Monitoring Ion Channel Function In Real Time Through Quantum Decoherence}

\author{L.T. Hall}
\affiliation{Center for Quantum Computer Technology, School of
Physics, University of Melbourne, VIC 3010, Australia}

\author{C.D. Hill}
\affiliation{Center for Quantum Computer Technology, School of
Physics, University of Melbourne, VIC 3010, Australia}

\author{J.H. Cole}
\affiliation{Center for Quantum Computer Technology, School of
Physics, University of Melbourne, VIC 3010,
Australia}
\affiliation{Institute f\"ur Theoretische
Festk\"orperphysik and DFG-Center for Functional Nanostructures
(CFN), Universit\"at Karlsruhe, 76128 Karlsruhe, Germany}

\author{B. St\"adler}

\affiliation{Centre for Nanoscience and Nanotechnology, Department
of Chemical and Biomolecular Engineering, The University of
Melbourne, Parkville, Victoria 3010, Australia}

\author{F. Caruso}

\affiliation{Centre for Nanoscience and Nanotechnology, Department
of Chemical and Biomolecular Engineering, The University of
Melbourne, Parkville, Victoria 3010, Australia}

\author {P. Mulvaney}

\affiliation{School of Chemistry and Bio21 Institute, University of
Melbourne, Parkville, Victoria 3010, Australia}

\author {J. Wrachtrup}

\affiliation{Physikalisches Institut, Universit\"at Stuttgart,
70550 Stuttgart, Germany}

\author{L.C.L. Hollenberg}
\affiliation{Center for Quantum Computer Technology, School of
Physics, University of Melbourne, VIC 3010, Australia}



 \maketitle
In drug discovery research there is a clear and urgent need for non-invasive detection of cell membrane ion channel operation with wide-field capability \cite{Lun06}. Existing techniques are generally invasive \cite{Hil01}, require specialized nano structures \cite{Fan02,Yam05,Rei08,Jel09}, or are only applicable to certain ion channel species \cite{Dem05}. We show that quantum nanotechnology has enormous potential to provide a novel solution to this problem. The nitrogen-vacancy (NV) centre in nano-diamond is currently of great interest as a novel single atom quantum probe for nanoscale processes \cite{Bal08,Neu07,Fu07,Cha07,Fak08,Bar09,Che04,Deg08,Tay08,Maz08,Col08,Bal09,Hal09}. However, until now, beyond the use of diamond nanocrystals as fluorescence markers \cite{Neu07,Fu07,Cha07,Fak08,Bar09}, nothing was known about the quantum behaviour of a NV probe in the complex room temperature extra-cellular environment. For the first time we explore in detail the quantum dynamics of a NV probe in proximity to the ion channel, lipid bilayer and surrounding aqueous environment. Our theoretical results indicate that real-time detection of ion channel
operation at millisecond resolution is possible by directly monitoring the quantum decoherence of the NV probe. With the potential to scan and scale-up to an array-based system this conclusion may have wide ranging implications for nanoscale biology and drug discovery.
\begin{figure}
\includegraphics[width=7cm]{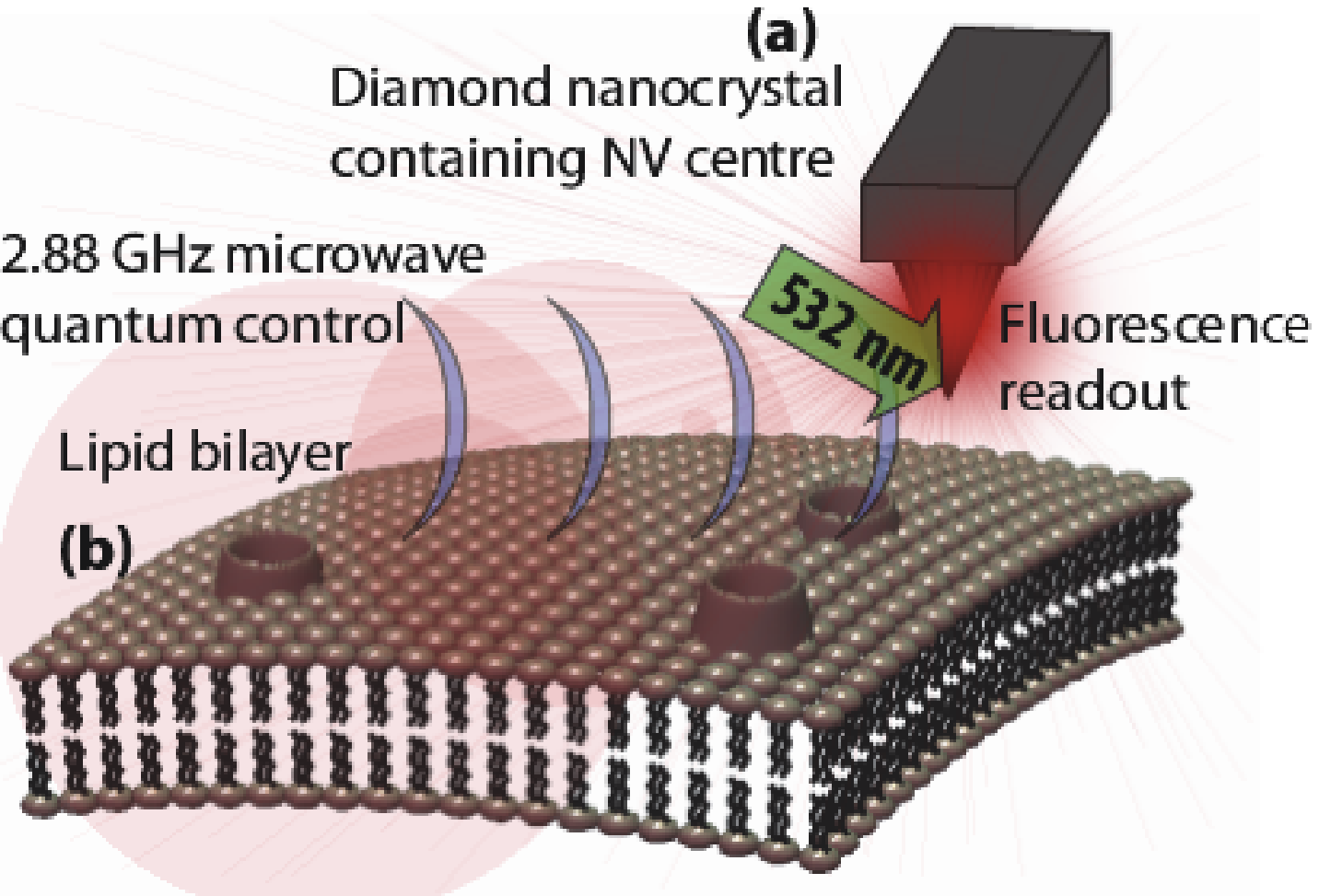}\\
\includegraphics[height=3.4cm]{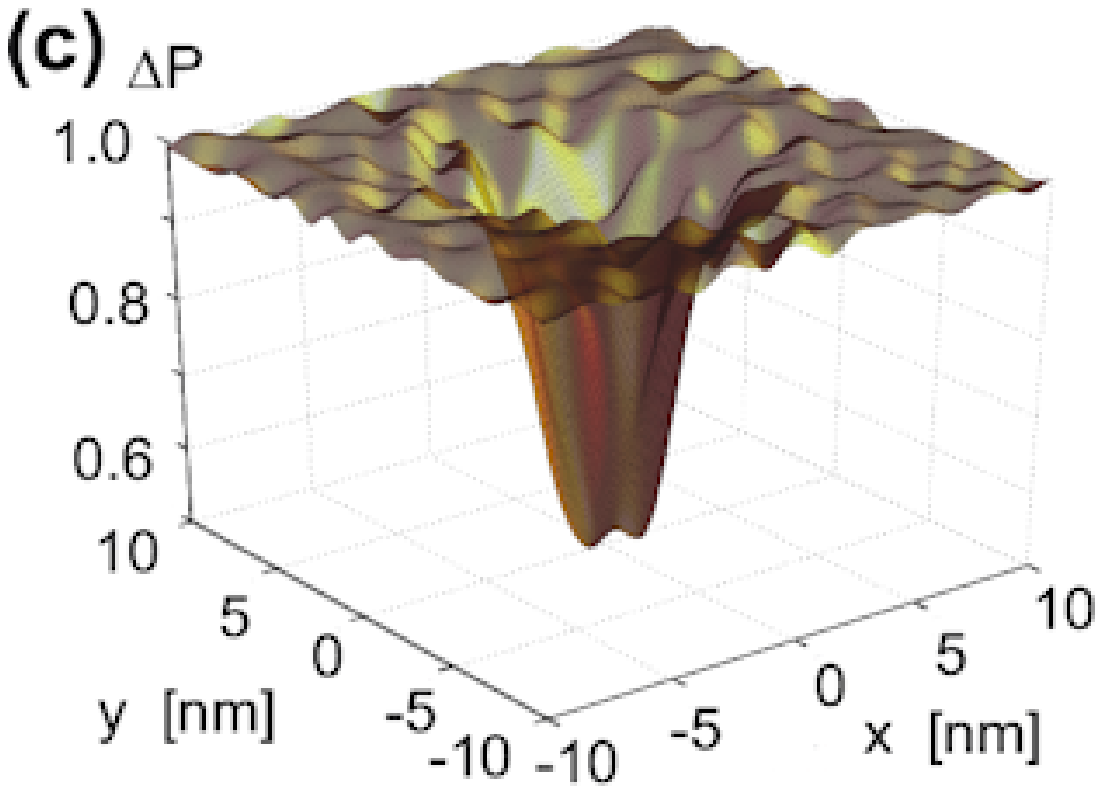}\includegraphics[height=3.2cm]{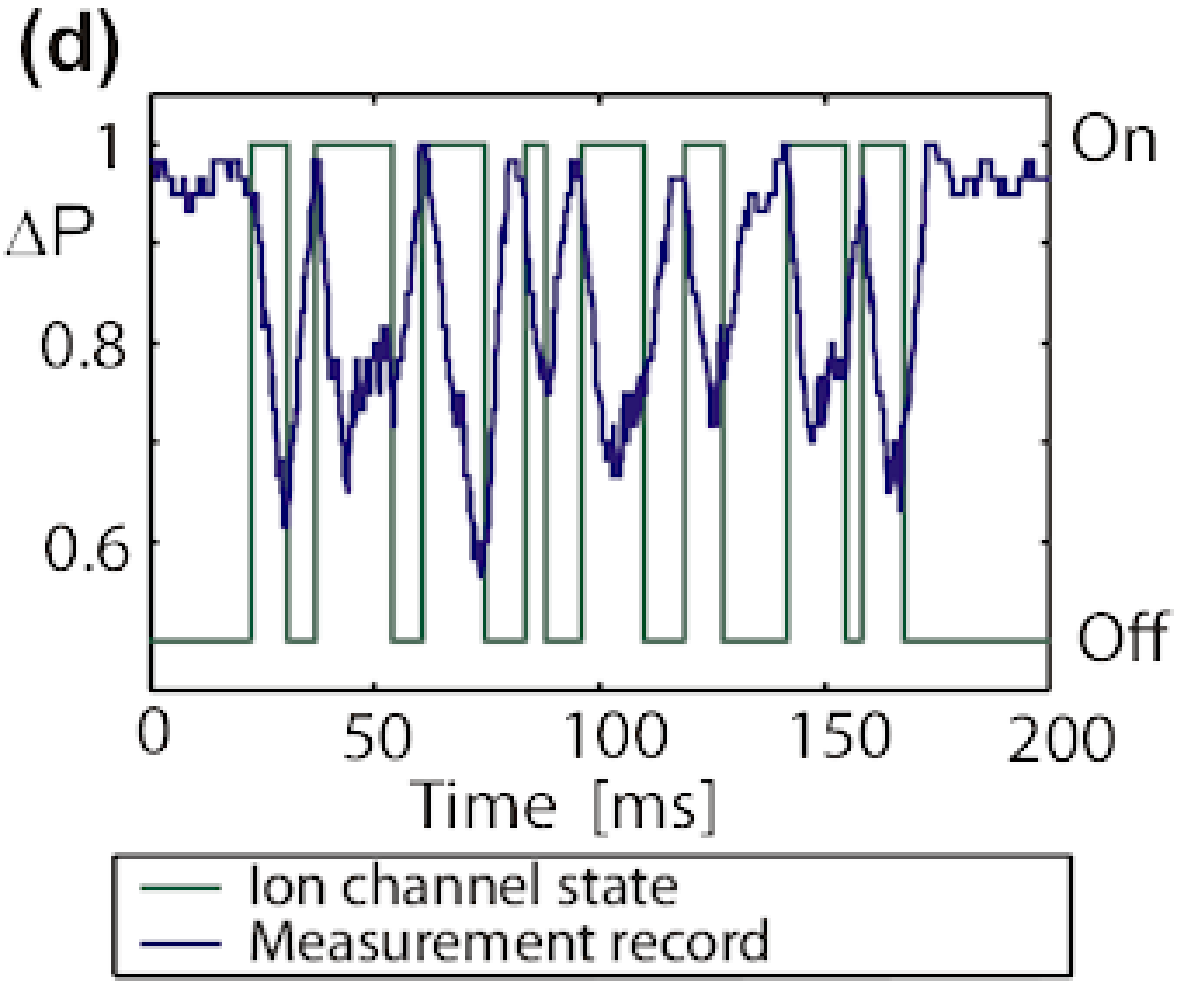}
\caption{Quantum decoherence imaging of ion channel operation. (a) A single nitrogen-vacancy (NV) defect in a diamond nanocrystal is placed on an AFM tip. The unique properties of the NV atomic level scheme allows for optically induced readout and microwave control of magnetic (spin) sub-levels. (b) The nearby cell membrane is host to channels permitting the flow of ions across the surface. The ion motion results in an effective fluctuating magnetic field at the NV position which decoheres the quantum state of the NV system. (c) This decoherence results in a decrease in fluorescence, which is most pronounced in regions close to the ion channel opening. (d) Changes in fluorescence also permit the temporal tracking of ion channel dynamics. } \label{device-fig}
\end{figure}

The cell membrane is a critical regulator of life. Its importance is reflected by the fact that the majority of drugs target membrane interactions \cite{Rei08}. Ion channels allow for passive and selective diffusion of ions across the cell membrane \cite{Ide02}, while ion pumps actively create and maintain the potential gradients across the membranes of living cells \cite{Baa08}. To monitor the effect of new drugs and drug delivery mechanisms a wide field ion channel monitoring capability is essential. However, there are significant challenges facing existing techniques stemming from the fact that membrane proteins, hosted in a lipid bilayer, require a complex environment to preserve their structural and functional integrity. Patch clamp techniques are generally invasive, quantitatively inaccurate, and difficult to scale up \cite{Dam05,Fen82,Qui02}, while black lipid membranes \cite{Mue62,Mue63} often suffer from stability issues and can only host a limited number of membrane proteins.

Instead of altering the way ion channels and the lipid membrane are presented or even assembled for detection, our approach is to consider a novel and inherently non-invasive in-situ detection method based on the quantum decoherence of a single-atom probe\cite{Col08}. In this context, decoherence refers to the loss of quantum coherence between magnetic sub-levels of a controlled atom system due to interactions with an environment. Such superpositions of quantum states are generally fleeting in nature due to interactions with the environment, and the degree and timescale over which such quantum coherence is lost can be measured precisely. The immediate consequence of the fragility of the quantum coherence phenomenon is that detecting the loss of quantum coherence (decoherence) in a single atom probe offers a unique monitor of biological function at the nanoscale.

The NV probe [Fig.\,\ref{device-fig}] consists of a nano-crystal
of diamond containing a nitrogen-vacancy (NV) defect placed at the end of
an AFM tip, as recently demonstrated \cite{Bal08}. For biological
applications a quantum probe must be submersible to be
brought within nanometers of
the sample structure, hence the NV system locked and protected in the ultra-stable diamond matrix [Fig.\,\ref{device-fig} (a)] is the system of choice. Of all the atomic systems known, the
NV centre in diamond alone offers the controllable, robust and persistent quantum properties such room temperature nano-sensing applications will demand \cite{Neu07,Fu07,Yu05}, as well as zero toxicity in a biological environment \cite{Yu05,Sch07,Bar09}.
Theoretical proposals for the use of diamond nanocrystals containing a NV system as sensitive nanoscale magnetometers \cite{Che04,Deg08,Tay08} have been followed closely by demonstrations in recent proof-of-principle experiments \cite{Maz08,Bal08,Bal09}. However, such nanoscale magnetometers employ only a fraction of the potential of the quantum resource at hand and do not have the sensitivity to detect the minute magnetic moment fluctuations associated with ion channel operation. In contrast, our results show that measuring the quantum decoherence of the NV induced by the ion flux provides a highly sensitive monitoring capability for the ion channel problem, well beyond the limits of magnetometer time-averaged field sensitivity \cite{Hal09}.

In order to determine the sensitivity of the NV probe to the ion channel signal we describe, for the first time, the lipid membrane, embedded ion channels, and the immediate surroundings as a fluctuating electromagnetic environment and quantitatively assess each effect on the quantum coherence of the NV centre. We consider the net magnetic field due to diffusion of nuclei, atoms and molecules in the immediate surroundings of the nanocrystal containing the NV system and the extent to which each source will decohere the quantum state of the NV. We find that, over and above these background sources, the decoherence of the NV spin levels is in fact highly sensitive to the particular signal due to the ion flux through a single ion channel. Our theoretical findings demonstrate the potential of this approach to revolutionize the way ion channels and potentially other membrane bound proteins or interacting species are characterized and measured, particularly when scale-up and scanning capabilities are considered.

\begin{figure}
\includegraphics[width=8.5cm]{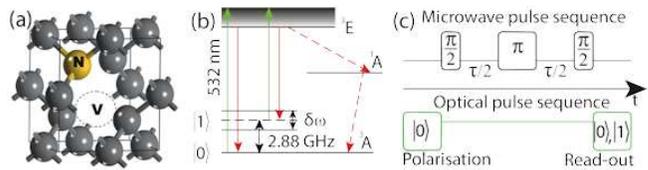}\\
\caption{(a) NV-centre diamond lattice defect. (b) NV spin
detection through optical excitation and emission cycle. Magnetic sublevels $m_s = 0$ and $m_s = \pm1$ are split by a $D$=2.88 GHz crystal field. Degeneracy between the $m_s = \pm1$ sublevels is lifted by a Zeeman shift, $\delta\omega$.
Application of 532 nm green light induces a spin-dependent photoluminescence and pumping into the $m_s = 0$ ground state.
(c) Microwave and optical pulse sequences for coherent control and readout.} \label{device-fig2}
\end{figure}

This paper is organized as follows. We begin by describing the
quantum decoherence imaging system [Fig.\,\ref{device-fig}] implemented using an NV centre in a realistic
technology platform. The biological system is described in detail,
considering the various sources of magnetic field fluctuations due
to atomic and molecular processes in the membrane itself and in the
surrounding media; and their effect on the decoherence of the optically monitored NV
system. Estimates of
the sensitivity of the NV decoherence to various magnetic field
fluctuation regimes (amplitude and frequency) are made which
indicate the ability to detect ion channel switch-on/off events.
Finally, we conduct large scale numerical simulations of the time
evolution of the NV spin system including all magnetic field
generating processes. This acts to verify the analytic picture, and provides
quantitative results for the monitoring and scanning capabilities of
the system.

The energy level scheme of the C$_{3v}$-symmetric NV system [Fig.\,\ref{device-fig2}(b)] consists of ground ($^3$A), excited ($^3$E) and
meta-stable ($^1$A) states. The ground state manifold has spin
sub-levels ($m=0,\pm1)$, which in zero field are split by 2.88 GHz.
In a background magnetic field the lowest two states ($m=0, +1$) are readily
accessible by microwave control. An important property of the NV
system is that under optical excitation the spin levels are
readily distinguishable by a difference in fluorescence, hence
spin-state readout is achieved by purely optical
means \cite{Jel02,Jel06}. Because of this relative simplicity of control
and readout, the quantum properties of the NV system, including the
interaction with the immediate crystalline environment, have been
well probed \cite{Jel04,Han08}. Remarkably for the decoherence imaging
application, the coherence time of the spin levels is very
long even at room temperature: in type 1b nanocrystals $T_2\sim 1
\mu s$, and in isotopically engineered diamond can be as
long as 1.8 ms\cite{Bal09} with the use of a spin-echo microwave control sequence [Fig.\,\ref{device-fig2}(c)].

Typical ion channel species K$^+$, Ca$^{2+}$, Na$^+$, and nearby water molecules are electron
spin paired, so any magnetic signal due to
ion channel operation will be primarily from the motion of nuclear
spins. Ions and water molecules enter the channel in thermal equilibrium with
random spin orientations, and move through the channel over a
$\mu$s timescale. The monitoring of ion channel activity occurs via measurement of the contrast in probe behavior between the on and off states of the ion channel. This then requires the dephasing due to ion channel activity to be at least comparable to that due to the fluctuating background magnetic signal.
We must therefore account for the decoherence of the NV quantum state due to the diffusion of water molecules,
buffer molecules, saline components as well as the transversal diffusion of lipid
molecules in the cell membrane.

The $n$th nuclear spin with charge $q_n$, gyromagnetic ratio $\gamma_n$, velocity $\vec{v}_n$
and spin vector $\vec{S}_n$, interacts with the NV spin
vector $\vec{P}$ and gyromagnetic ratio $\gamma_{\rm p}$ through the
time-dependent dipole dominated interaction:
\begin{eqnarray}
H_{\rm int} (t) &=& \sum_{n=1}^{N} \kappa^{(n)}_{\mathrm{dip}}  \left[ {\vec{P}\cdot
\vec{S}_n \over r^3_n(t)} - 3{ \vec{P}\cdot \vec{r}_n(t) \vec{S}_n
\cdot \vec{r}_n(t) \over r^3_n(t)}\right]
\end{eqnarray} where $\kappa^{(n)}_{\mathrm{dip}} \equiv {\mu_0 \over 4 \pi}\hbar^2
\gamma_{\rm p} \gamma_n $ are the probe-ion coupling strengths,
and $\vec{r}_n(t)$ is the time-dependent ion-probe separation. Additional Biot-Savart fields generated by the ion motion, both in the channel and the extracellular environment, are
several orders of magnitude smaller than this dipole interaction and
are neglected here. Any macroscopic fields due to intracellular ion currents are of nano-Tesla (nT) order and are effectively static over $T_2$ timescales. These effects will thus be suppressed by the spin-echo pulse sequence.

\begin{figure}
\includegraphics[width=8cm]{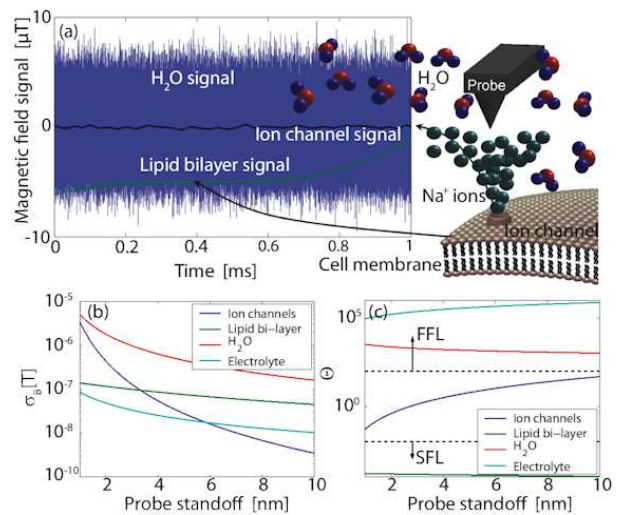}\\
\caption{(a) Typical magnetic field signals from water, ion channel and lipid bilayer sources at a probe standoff of 4\,nm over a 1\,ms timescale. (b) Comparison of $\sigma_B$ for various sources of magnetic fields. (c) Fluctuation regime, $\Theta = f_\mathrm{e}/\gamma_\mathrm{p}\sigma_\mathrm{B}$, for magnetic field sources vs probe standoff. Rapidly fluctuating fields ($\Theta\gg1$) are said to be in the fast-fluctuating limit (FFL). Slowly fluctuating fields ($\Theta\ll1$) are in the slow fluctuation limit (SFL). The ion channel signal exists in the $\Theta\sim1$ regime, and therefore has an optimal dephasing effect on the NV probe.
}
\label{gspop}
\end{figure}

In Fig.\,\ref{gspop}(a) we show typical field traces at a probe height of 1-10nm above the ion channel, generated by the ambient environment and the on-set of ion-flow as the channel opens. The contribution to the net field at the NV probe position from the various background diffusion processes dominate the ion channel signal in terms of their amplitude. Critically, since the magnetometer mode
detects the field by acquiring phase over the coherence time of the
NV centre, both the ion channel signal and background are well below
the nT\,Hz$^{-1/2}$ sensitivity limit of the magnetometer over the ($\sim 1$\,ns) self-correlated timescales of the environment. However, the effect of
the various sources on the decoherence rate of the NV centre are
distinguishable because the amplitude-fluctuation frequency scales
are very different, leading to remarkably different dephasing behaviour.

To understand this effect, we need to consider the full quantum evolution of the
NV probe. In the midst of this environment the probe's quantum
state, described by the density matrix $\rho(t)$, evolves according
to the Liouville equation,
${d \over dt}\rho(t) = -{i\over \hbar} \bigl[H(t)\rho(t)-\rho(t)H(t)\bigr]$,
where $\rho(t)$ is the incoherent thermal average over all possible unitary evolutions of the entire system, as described by the full Hamiltonian, $H = H_{\mathrm{nv}}+H_\mathrm{int}+H_\mathrm{bg},$
where $H_{\rm nv}$ is the Hamiltonian of the NV system, and $H_{\rm int}$ describes the interaction of the NV system with the background environment (e.g. diffusion of ortho spin water
species and ions in solution) and any intrinsic coupling
to the local crystal environment (e.g. due to $^{13}$C nuclei or
interface effects).
The evolution of the background system due to self interaction is described by $H_\mathrm{bg}$, which, in the present methodology, is used to obtain the noise spectra of the various background processes. We note that the following analysis assumes dephasing to be the dominant decoherence channel in the system. We ignore relaxation processes since all magnetic fields considered are at least 4 orders of magnitude less than the effective crystal field of $D/\gamma_\mathrm{p}\sim0.2$\,T, and are hence unable to flip the probe spin over relevant timescales. Phonon excitation in the diamond crystal leads to relaxation times of the order of 100\,s \cite{Bal09} and may also be ignored. Before moving
onto the numerical simulations we consider some important features
of the problem.

The decoherence rate of the NV centre is governed by the accumulated
phase variance during the control cycle. Maximal dephasing due to a fluctuating field will occur at the cross-over point between the fast (FFL) and slow (SFL) fluctuation regimes \cite{Hal09}. A
measure of this cross-over point is the dimensionless ratio $\Theta \equiv f_{\rm e}/\gamma_{\rm p}
\sigma_B$, where $\tau_\mathrm{e}=1/f_\mathrm{e}$ is the correlation time of the fluctuating signal, with cross-over at $\Theta \sim 1$. We can estimate the
field standard deviation $\sigma_B^{\rm ic}$ due to the random
nuclear spin of ions and bound water molecules moving in an ion channel (ic) as:
\begin{eqnarray}
\sigma_B^{\rm ic}&\sim&\frac{\mu_0}{4\pi}\frac{1}{h_p^{3}}\sqrt{N_\mathrm{ion}\mu_\mathrm{ion}^2+N_\mathrm{H_2O}\mu_\mathrm{H_2O}^2}.
\end{eqnarray} The fluctuation strength of the ion channel magnetic field, $\sigma_B^{\rm ic}$, is plotted in Fig.\,\ref{gspop}(b) as a function of the probe stand-off distance, $h_p$. Ion flux rates are of the order of $\sim5\times10^{-4}\mathrm{\,\,ions\,\,ns^{-1}nm^{-2}}$\cite{Leo07}, giving an effective dipole field fluctuation rate of $f_e\sim 3\times10^4\mathrm{\,\,Hz}$. For probe-channel separations of 2-8nm, values of $\Theta$ range from 0.4 to 40 [Fig.\,\ref{gspop}(c)]. Thus, the ion channel flow
hovers near the cross-over point, with an induced dephasing rate of $\Gamma_\mathrm{ic}\sim10^4-10^5\mathrm{\,\,Hz}$.

We now consider the dephasing effects of the various sources of background magnetic fields. The first source of background noise is the fluctuating magnetic field arising from the motion of the water molecules and ions throughout the aqueous solution. Due to the nuclear spins of the hydrogen atoms, liquid water consists of a mixture of spin neutral (para) and spin-1 (ortho) molecules. The equilibrium ratio of ortho to para molecules (OP ratio) is 3:1 \cite{Tik02}, making 75\% of water molecules magnetically active. In biological conditions, dissolved ions occur in concentrations 2-3 orders of magnitude below this and are ignored here (they are important however for calculations of the induced Stark shift, see below). The RMS strength of the field due to the aqueous solution is
\begin{eqnarray}
  \sigma_B^{\mathrm{H_2O}}&\sim& g_\mathrm{H}\mu_\mathrm{N}\frac{\mu_0}{2\pi}\sqrt{ n_{\mathrm{H_2O}} \frac{\pi}{h_p^3}}.
\end{eqnarray} This magnetic field is therefore 1-2 orders of magnitude stronger than the field from the ion channel [Fig\, \ref{gspop}(a,b)].
The fluctuation rate of the aqueous environment is dependent on the self diffusion rate of the water molecules. Using $D_{\mathrm{H_2O}}=3\times10^{-9}\mathrm{\,\,m^2\,s^{-1}}$, the fluctuation rate is $f_e^{\mathrm{H_2O}}\sim D_{\mathrm{H_2O}}/\left(2h_\mathrm{p}\right)^2$. This places the magnetic field due to the aqueous solution in the fast-fluctuation regime, with $\Theta_{\mathrm{H_2O}}\sim10^3-10^4$ [Fig.\,\ref{gspop}(b)], giving a comparatively slow dephasing rate of $\Gamma_{\mathrm{H_2O}}\sim f_e^{\mathrm{H_2O}}\Theta_{\mathrm{H_2O}}^{-2}\sim100\mathrm{\,\,Hz}$ and corresponding dephasing envelope $\mathcal{D}_\mathrm{H_2O} = e^{-\Gamma_{\mathrm{H_2O}}t}$.

An additional source of background dephasing is the lipid molecules comprising the cell membrane. Assuming magnetic contributions from hydrogen nuclei in the lipid molecules, lateral diffusion in the cell membrane gives rise to a fluctuating B-field, with a characteristic frequency related to the diffusion rate. Atomic hydrogen densities in the membrane are $n_\mathrm{H}
\sim 3\times 10^{28}\,\mathrm{m}^{-3} $.
At room temperature, the populations of the spin states of
hydrogen will be equal, thus the RMS field strength is given by
\begin{eqnarray}
  \sigma_B^{\mathrm{L}} &\sim& g_H\mu_N\frac{\mu_0}{8\pi}\sqrt{n\frac{5\pi }{4h_\mathrm{p}^3}}.
\end{eqnarray} The strength of the fluctuating field due to the lipid bilayer is of the order of $10^{-7}$\,T [Fig.\,\ref{gspop}(a)]. The Diffusion constant for lateral Brownian motion of lipid
molecules in lipid bilayers is $D_\mathrm{L} =
2\times10^{-15}\,\mathrm{m^2s^{-1}}$ \cite{Ban06}, giving a fluctuation frequency of $f_e^{\mathrm{L}}\sim125 \mathrm{\,\,Hz}$ and $\Theta_\mathrm{L}\sim10^{-4}$ [Fig.\,\ref{gspop}(d)]. At this frequency, any quasi-static field effects will be predominantly suppressed by the spin-echo refocusing. The leading-order (gradient-channel) dephasing rate is given by \cite{Hal09},
\begin{eqnarray}
  \Gamma_\mathrm{L} &\sim& \frac{1}{2\sqrt{2\sqrt2}}\Theta_\mathrm{L}^{-1/2}f_e^\mathrm{L} +\mathcal{O}\biggl(\Theta_\mathrm{L}^{-1/3}f_e^\mathrm{L}\biggr),
\end{eqnarray} giving rise to dephasing rates of the order $\Gamma_\mathrm{L}\sim100$\,Hz, with corresponding dephasing envelope $\mathcal{D}_\mathrm{L}(t) = e^{-\Gamma^4_\mathrm{L}t^4}$.

The electric fields associated with the dissolved ions also interact with the NV centre via the ground state Stark effect. The coefficient for the frequency shift as a function of the electric field applied along the dominant ($z$) axis is given by $R_{3D} = 3.5\times10^{-3}\mathrm{\,\,Hz\,m\,V^{-1}}$ \cite{Van90}. Fluctuations in the electric field may be related to an effective magnetic field via $B_z^\mathrm{eff} = R_{3D}E_z/\gamma_p$, which may be used in an analysis similar to that above.
An analysis using Debye-H\"uckel theory \cite{Kim08} shows charge fluctuations of an ionic solutions in a spherical region $\Lambda$ of radius $R$ behave as
\begin{equation}
  \langle Q^2_\Lambda\rangle = D_\mathrm{E}k_\mathrm{B}T\left(1+\kappa R\right)e^{-\kappa R}\left[R\cosh\left(\kappa R\right) - \frac{\sinh\left(\kappa R\right)}{\kappa}\right]\label{Qsq},
\end{equation} where $D_\mathrm{E}$ is the diffusion coefficient of the electrolyte, and $\kappa$ is the inverse Debye length ($l_\mathrm{D}$); $l_\mathrm{D} = 1/\kappa = 1.3$ nm for biological conditions. Whilst this analysis applies to a region $\Lambda$ embedded in an infinite bulk electrolyte system, simulation results discussed below show very good agreement when applied to the system considered here. Eq.\,\ref{Qsq} is used to obtain the electric field variance, $\sigma_\mathrm{E} = \sqrt{\langle E^2 \rangle - \langle E \rangle^2}\sim10^6\,\mathrm{Vm^{-1}}$, as a function of $h_\mathrm{p}$. Relaxation times for electric field fluctuations are $\tau_e^\mathrm{E} = \epsilon\epsilon_0\rho_\mathrm{E}$ \cite{For00}, where $\rho_\mathrm{E}$ is the resistivity of the electrolyte, giving $f_e^\mathrm{E} \sim 1/\tau_e^\mathrm{E} = 1.4\times10^9\mathrm{\,\,Hz}$ under biological conditions. Given the relatively low strength [Fig.\,\ref{gspop}(a)] and short relaxation time of the effective Stark induced magnetic field fluctuations ($\Theta\sim10^5$) [Fig.\,\ref{gspop}(b)], we expect the charge fluctuations associated with ions in solution to have little effect on the evolution of the probe.

\begin{figure}
\includegraphics[width=8.5cm]{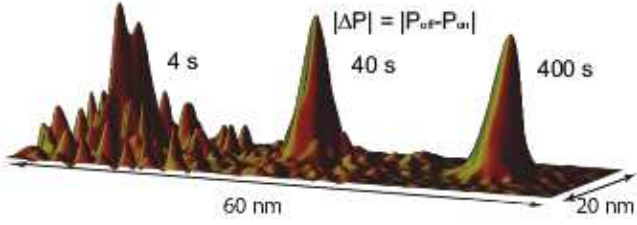}
\caption{Spatial information corresponding to the ion channel as a dephasing source. Relative population differences are plotted for pixel dwell times of 10, 100 and 1000 ms. Corresponding image acquisition times are 4, 40 and 400 s.}
\label{scans}
\end{figure}

We now turn to the problem of non-invasively resolving the location of a sodium ion channel in a lipid bilayer membrane. When the channel is closed, the dephasing is the result of the background activity, and is defined by $\mathcal{D}_\mathrm{off}=\mathcal{D}_\mathrm{H_2O}\mathcal{D}_\mathrm{L}\mathcal{D}_\mathrm{E}\mathcal{D}_\mathrm{^{13}C}$. When the channel is open, the dephasing envelope is defined by $\mathcal{D}_\mathrm{on}=\mathcal{D}_\mathrm{off}\mathcal{D}_\mathrm{ic}$.  Maximum contrast will be achieved by optimising the spin-echo interrogation time, $\tau$, to ensure $\mathcal{D}_\mathrm{off} - \mathcal{D}_\mathrm{on}$ is maximal. Thus in the vicinity of an open channel at the point of optimal contrast, $\tau \approx T_2/2$, we expect an ensemble ground state population of $P_\mathrm{on}(\frac{T_2}{2}) = \frac{1}{2}\left[1+\mathcal{D}_\mathrm{on}\left(\frac{T_2}{2}\right)\right]=0.61$, and $P_\mathrm{off}(\frac{T_2}{2}) = \frac{1}{2}\left[1+\mathcal{D}_\mathrm{off}\left(\frac{T_2}{2}\right)\right]=0.93$ otherwise. By scanning over an open ion channel and monitoring the probe via repeated measurements of the spin state, we may build up a population ensemble for each lateral point in the sample. The signal to noise ratio improves with the dwell time at each point. Fig. \ref{scans} shows simulated scans of a sodium ion channel with corresponding image acquisition times of 4, 40 and 400\,s. It should be noted here that the spatial resolution available with this technique is beyond that achievable by magnetic field measurements alone, since for large $\Theta$, $\Delta P \propto B^2 \propto h_p^{-6}$.

We may employ similar techniques to temporally resolve a sodium ion channel switch-on event. By monitoring a single point, we may build up a measurement record sequence, $\mathcal{I}$.
In an experimental situation, the frequency with which measurements may be performed has an upper limit of $f_\mathrm{m}=\left(\tau+\tau_\mathrm{m}+\tau_{2\pi}\right)^{-1}$, where 
$\tau_\mathrm{m}\approx900$ ns is the time required for photon collection, and $\tau_{2\pi}$ is the time required for all 3 microwave pulses. A potential trade-off exists between the increased dephasing due to longer interrogation times and the corresponding reduction in measurement frequency. Interrogation times are ultimately limited by the intrinsic $T_2$ time of the crystal. A second trade-off exists between the variance of a given set of $N_\tau$ consecutive measurements and the temporal resolution of the probe. For the monitoring of a switching event, the spin state population may be inferred with increased confidence by performing a running average over a larger number of data points, $N_\tau$. However increasing $N_\tau$ will lead to a longer time lag before a definitive result is obtained. The uncertainty in the ion channel state goes as $\delta P \sim \left(\sqrt{N_\tau}\right)^{-1}$, where $N_\mathrm{\tau}$ is the number of points included in the dynamic averaging. We must take sufficient $N_\mathrm{\tau}$ to ensure that $\delta P < \Delta P(\tau,h_\mathrm{p},T_2) = P_\mathrm{off} - P_\mathrm{on}$. The temporal resolution depends on the width of the dynamic average and is given by $\delta t\sim N_\mathrm{\tau}(\tau+\tau_m)$, giving the relationship
\begin{eqnarray}
  \delta t &=& \frac{\tau +\tau_m}{\delta P^2}\,\, >\,\, \frac{\tau +\tau_m}{\,\left[\Delta P\left(\tau,h_\mathrm{p},T_2\right)\right]^2}.
\end{eqnarray} We wish to minimise this function with respect to $\tau$ for a given stand-off ($h_\mathrm{p}$) and crystal $T_2$ time.
\begin{figure}
\includegraphics[width=8.4cm]{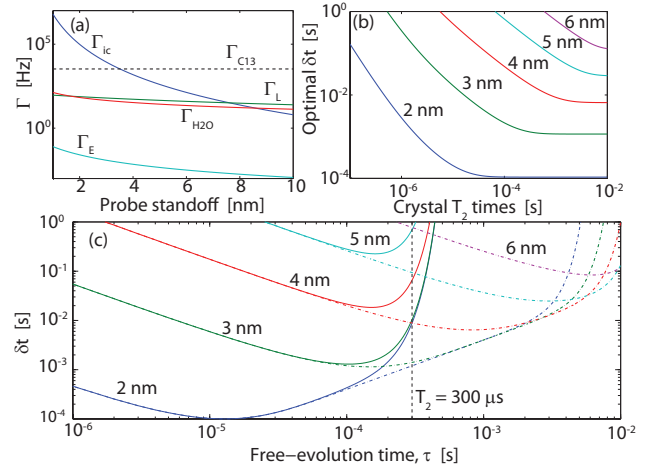}\\
\caption{(a) Dephasing rates due to the sources of magnetic field plotted as a function of probe standoff, $h_p$. (b) Optimum temporal resolution as a function of crystal $T_2$ times for $h_p = 2-6$ nm. (c) Temporal resolution as a function of interrogation time, $\tau$, for separations of 2 - 6 nm and $T_2 = 300\,\mu$s. Dashed lines show expected improvements from much longer $T_2$ times, $T_2\gg\tau$.} \label{switchpop}
\end{figure}

In reality, not all crystals are manufactured with equal $T_2$ times. An important question is therefore, for a given $T_2$, what is the best temporal resolution we may hope to achieve? Fig.\,\ref{switchpop}(b) shows the optimal temporal resolution as a function of $T_2$. It can be seen that $\delta t$ improves monotonically with $T_2$ until $T_2$ exceeds the dephasing time due the fluctuating background fields [Fig.\,\ref{switchpop}(a)]. Beyond this point no advantage is found from extending $T_2$.

A plot of $\delta t$ as a function of $\tau$ is given in Fig.\,\ref{switchpop}(c) for standoffs of 2-6 nm. Solid lines depict the resolution that maybe achieved with $T_2=300\,\mu$s. Dashed lines represent the resolution that may be achieved by extending $T_2$ beyond the dephasing times of background fields. We see that $\delta t$ diverges as $\tau \rightarrow T_2$, and is optimal for $\tau \rightarrow 1/\Gamma_\mathrm{ic}$.

\begin{figure}
\includegraphics[width=8cm]{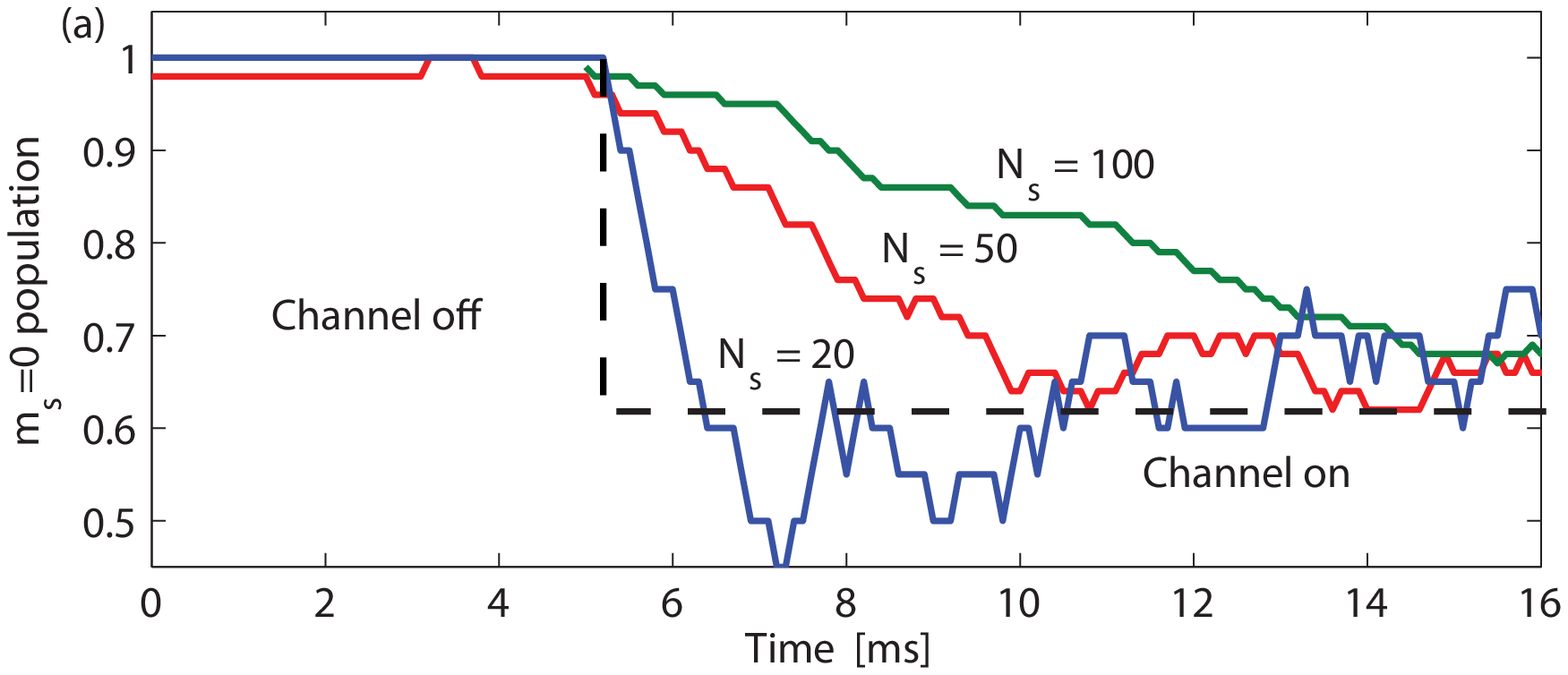}\\
\includegraphics[width=8cm]{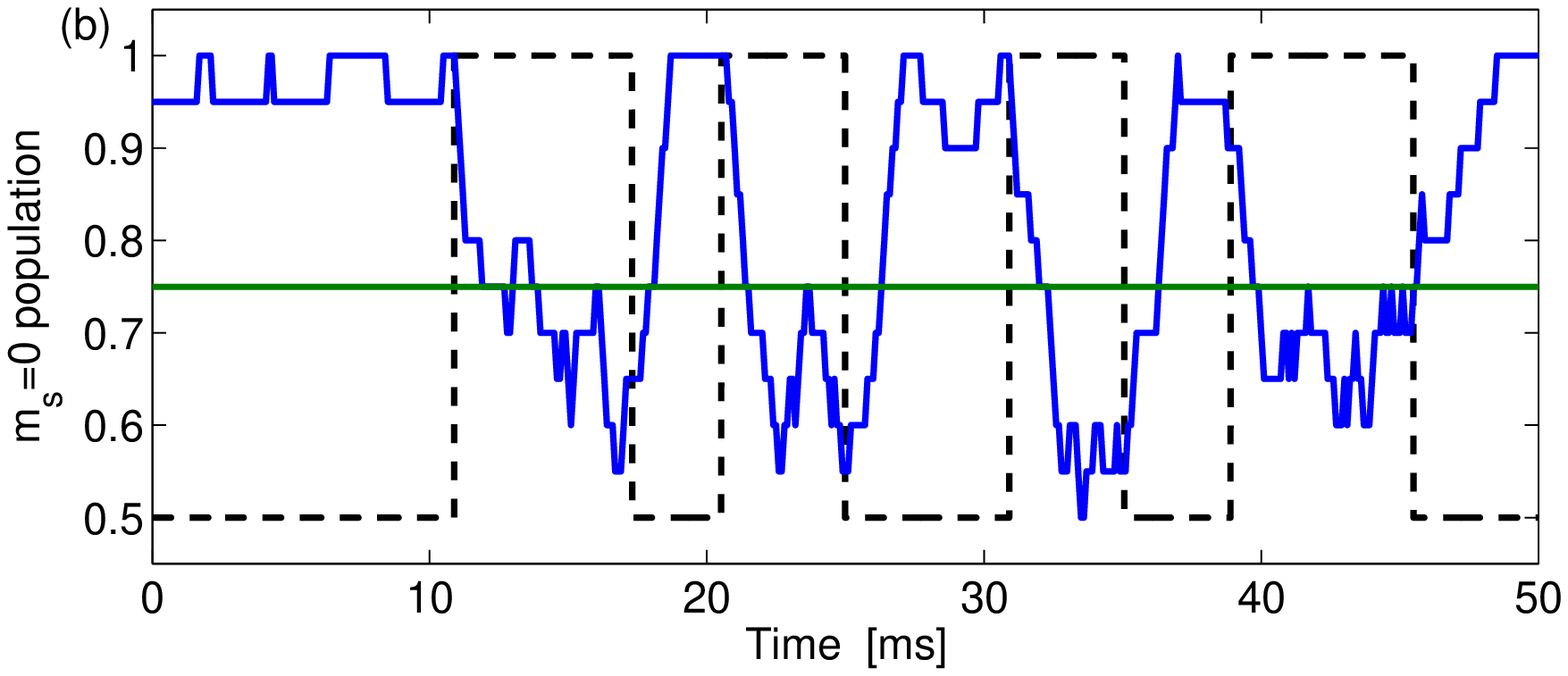}\\
\includegraphics[width=2.6cm]{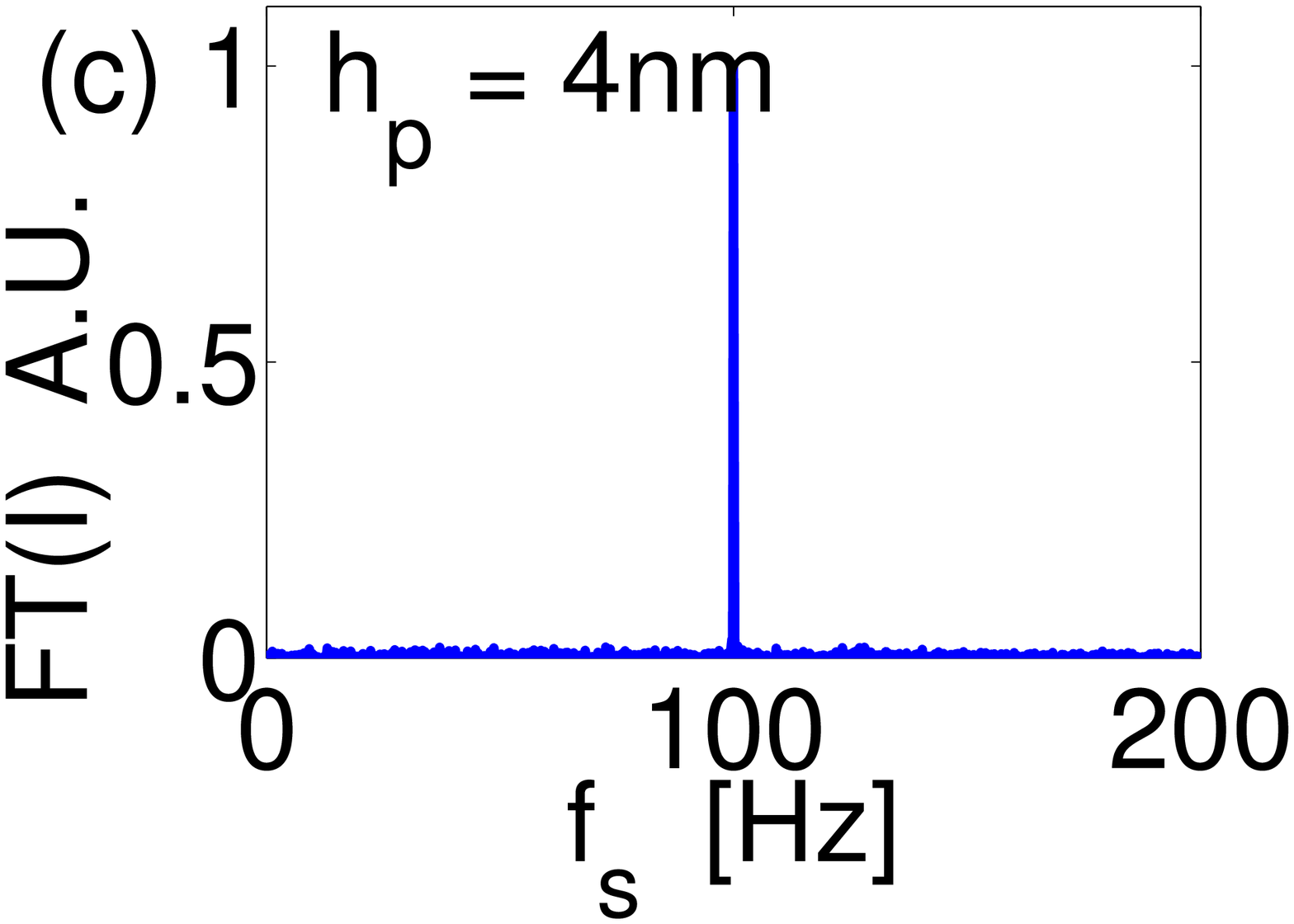}\includegraphics[width=2.6cm]{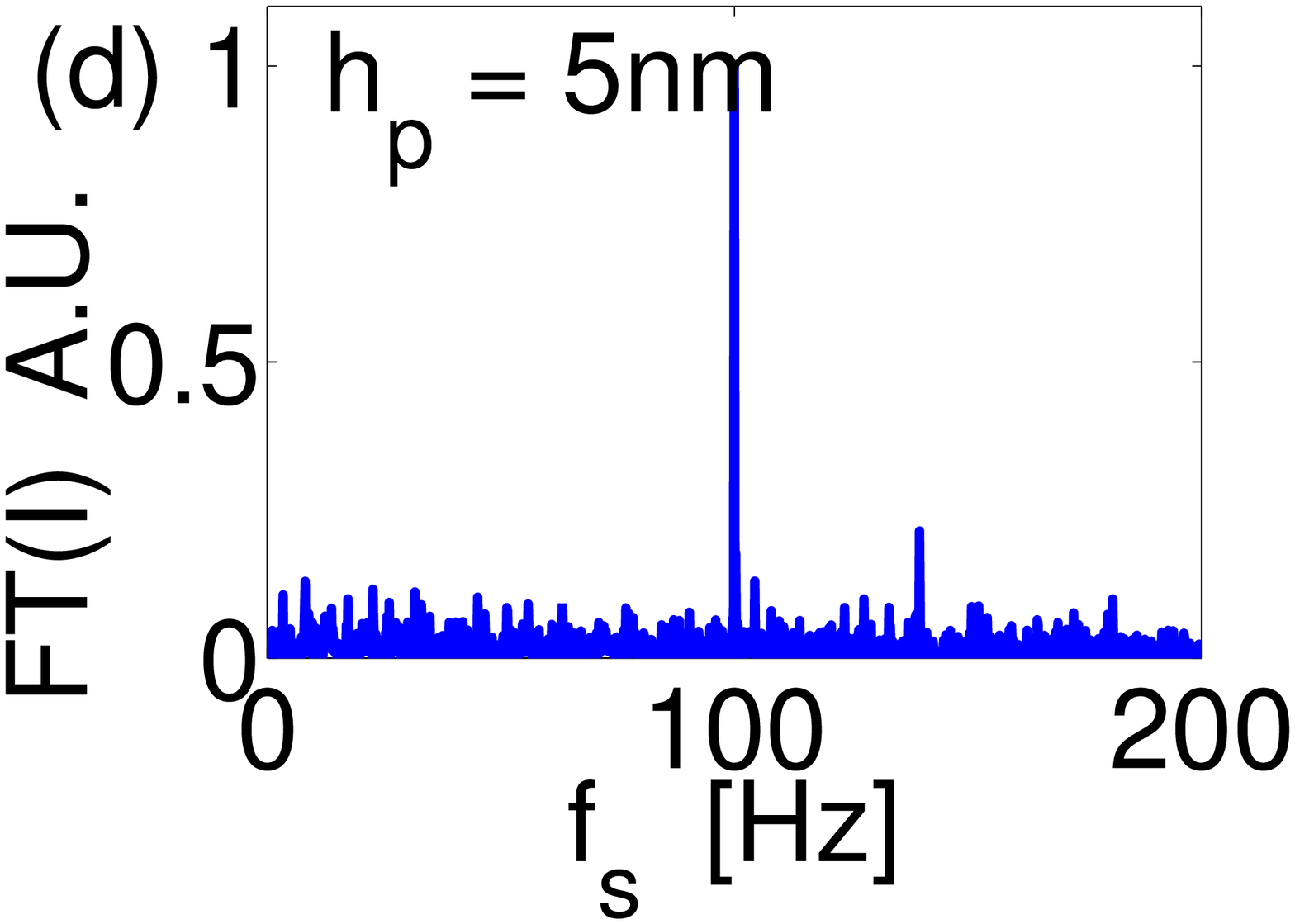}\includegraphics[width=2.6cm]{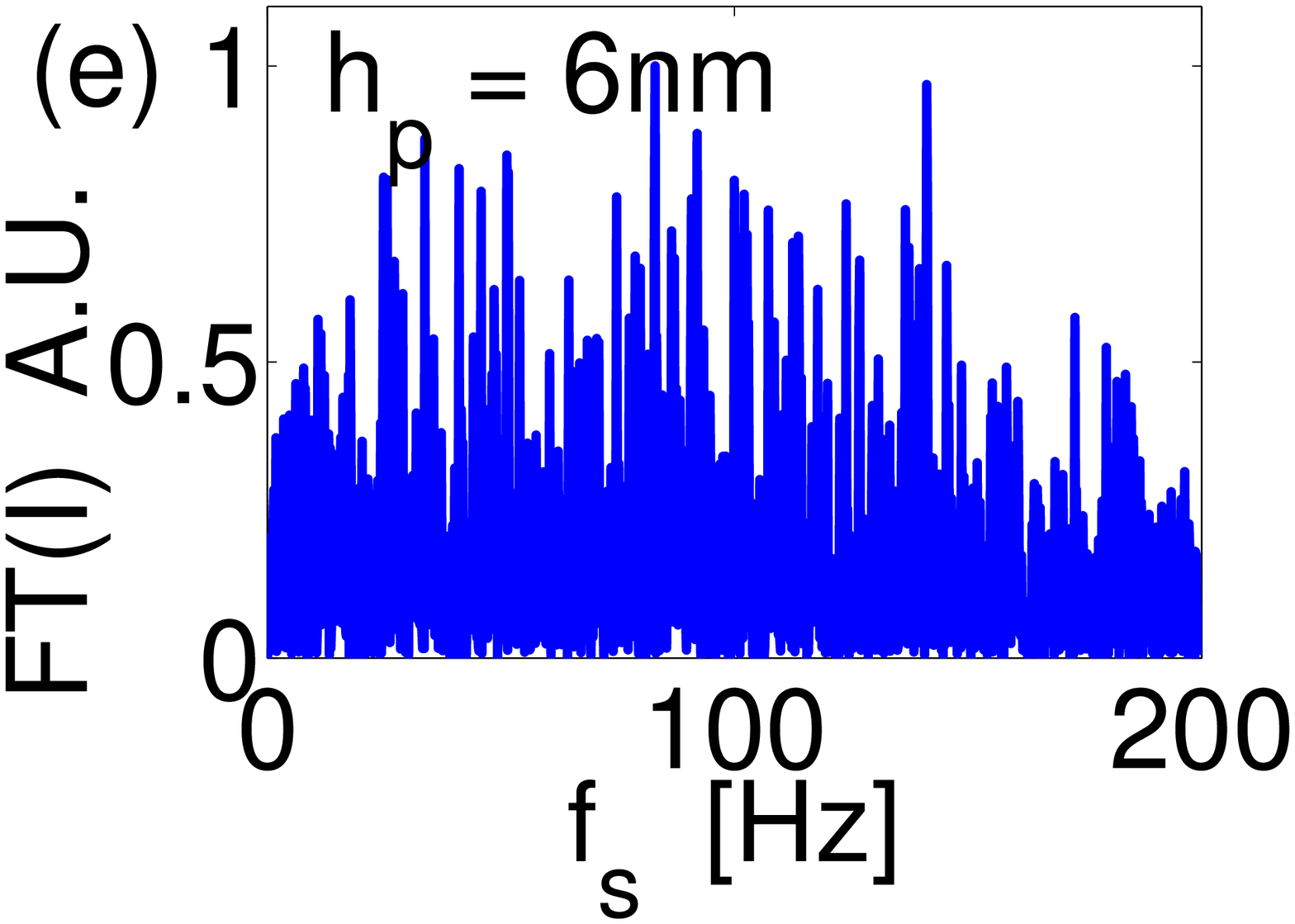}\\
\caption{(a) Plot illustrating the dependence of temporal resolution ($\delta t$) and signal variance ($\delta P$) on the number of data points included in the running average ($N_\mathrm{s}$). (b) Simulated reconstruction of a sodium ion channel signal with a 200 Hz switching rate using optical readout of an NV centre (blue curve). The actual ion channel state (on/off) is depicted by the dashed line, and the green line depicts the analytic confidence threshold. Fourier transforms of measurement records are shown in (c)-(e) for stand-offs of 4, 5 and 6 nm respectively. Switching dynamics are clearly resolvable for $h_p<6$ nm, beyond which there is little contrast between decoherence due to the ion channel signal and the background.}
\label{switch}
\end{figure}

As an example of monitoring of ion channel behaviour, we consider a crystal with a $T_2$ time of 300 $\mu$s at a standoff of 3 nm. Fig.\,\ref{switchpop}(c) tells us that an optimal temporal resolution of $\delta t \sim 1.1$ ms may be achieved by choosing $\tau\sim100\,\mu$s. This in turn suggests an optimal running average will employ $N_\mathrm{\tau} = \delta t\left(\tau+\tau_\mathrm{m}\right)^{-1}\approx$ 11 data points. Fig.\,\ref{switch}(a) shows a simulated detection of a sodium ion channel switch-on event using $N_\mathrm{\tau}=20,\,\,50$ and 100 points. The effect of increasing $N_\mathrm{\tau}$ is shown to give poorer temporal resolution but also produces a lower variance in the signal. This may be necessary if there is little contrast between $P_\mathrm{off}$ and $P_\mathrm{on}$. Conversely, decreasing $N_\mathrm{\tau}$ results in an improvement to the temporal resolution but leads to a larger signal variation.

We now consider an ion channel switching between states after an average waiting time of 5 ms (200 Hz) [Fig.\,\ref{switch}(b)]. To ensure the condition $\delta P < \Delta P$ is satisfied, we perform the analysis using $N_\mathrm{\tau}=20$, giving a resolution of $\delta t\approx2$ ms. The blue curve shows the response of the NV population to changes in the ion channel state. Fourier transforms of the measurement record, $\mathcal{F}\left(\mathcal{I}\right)$, are shown in Fig.\,\ref{switch}(c)-(e). The switching dynamics are clearly resolvable for heights less than 6 nm. The dominant spectral frequency is 100 Hz which is half the 200 Hz switching rate as expected. Beyond 6 nm, the contrast between $P_\mathrm{off}$ and $P_\mathrm{on}$ is too small to be resolvable due to the $T_2$ limited temporal resolution, as given in Fig.\,\ref{switchpop}(b). This may be improved via the manufacturing of nanocrystals with improved $T_2$ times, allowing for longer interrogation times [dashed curves, Fig.\,\ref{switchpop}(c)].


With regard to scale-up to a wide field imaging capability, beyond the obvious extrinsic scaling of the number of single channel detection elements (in conjunction with micro-confocal arrays), we consider an intrinsic scale-up strategy using many NV centres in a bulk diamond probe, with photons collected in a pixel arrangement. Since the activity of adjacent ion channels is correlated by the $\mu$m scale activity of the membrane, the fluorescence of adjacent NV centres will likewise be correlated, thus wide field detection will occur via a fluorescence contrast across the pixel. Implementation of this
scheme involves a random distribution of NV
centres in a bulk diamond crystal. The highest reported NV density is $2.8\times10^{24}$ m$^{-3}$ \cite{Aco09}, giving typical NV-
NV couplings of $<10$ MHz which are strong enough to introduce significant additional decoherence. We seek a compromise between increased population contrast and increased decoherence rates due to higher NV densities, $n_\mathrm{nv}$, given by $\Gamma_\mathrm{nv}\sim\frac{\sqrt{2\pi}}{3}\frac{\hbar\mu_0}{4\pi}\gamma_\mathrm{p}^2n_\mathrm{nv}$\cite{Hal09}.

For ion channel operation correlated across each pixel, the total population contrast $\Delta \Phi$ between off and on states is obtained by averaging the local NV state population change
$\Delta \Phi(\tau) = P_{\rm off}({\vec r}_i,{\vec r}_c,\tau) - P_{\rm on}({{\vec r}_i},{{\vec r}_c},\tau)$ over all NV positions ${\vec r}_i$ and orientations; and ion channel positions ${\vec r}_c$ and species; and maximizing with respect to $\tau$. As an example, consider a crystal with $n_\mathrm{nv}=10^{24}\mathrm{\,m^{-3}}$ whose surface is brought within 3 nm of the cell membrane containing an sodium and potassium ion channel densities of $\sim2\times10^{15}\,\mathrm{m}^{-2}$\cite{Arh07}. Higher densities will yield better results, however these have not been realised experimentally as yet, and electron spins in residual nitrogen will begin to induce NV spin flips. We expect ion channel activity to be correlated across pixel areas of 1\,$\mu$m $\times$ 1\,$\mu$m, so the population contrast between off and on states is $\Delta\Phi\approx15$. At these densities, the optimal interrogation time is $\tau\sim0.8\,\mu$s, yielding an improvement in the temporal resolution by a factor of 10,000, opening up the potential for single-shot measurements of ion channel activity across each pixel.

We have carried out an extensive analysis of the quantum dynamics of a NV diamond probe in the
cell-membrane environment and determined the theoretical sensitivity for the detection, monitoring and imaging
of single ion channel function through quantum decoherence. Using current demonstrated technology a temporal
resolution in the 1-10 ms range is possible, with spatial
resolution at the nanometer level. With the scope for
scale-up and novel scanning modes, this fundamentally
new detection mode has the potential to revolutionize
the characterization of ion channel action, and possibly
other membrane proteins, with important implications
for molecular biology and drug discovery.

\begin{acknowledgments}
This work was supported by the Australian Research Council.
\end{acknowledgments}


\end{document}